\documentstyle[psfig]{mn}
\def\refindent{\par \noindent \hang}
\def\paper#1#2#3#4#5{\refindent #1, #2, #3, #4, #5}
\def\press#1#2#3{\refindent #1, #2, #3, in press}
\def\conf#1#2#3#4#5#6{\refindent #1, #2, in #3, eds, #4. #5, p.~#6}
\def\preprint#1#2{\refindent #1, #2, Preprint}
\def\and{, }
\def\AaA{A\&A}
\def\AJ{AJ}

\def\ApJ{ApJ}
\def\ApJL{ApJ}
\def\ApJS{ApJS}

\def\ARAA{ARA\&A}

\def\JCP{J.~Comp.~Phys.}
\def\MN{MNRAS}

\def\etal{{\rm et al.\thinspace}}
\def\vec#1{{\bf #1}}
\def\dd{{\rm d}}
\def\hmpc{$h^{-1} {\rm Mpc}$}
\def\hkpc{$h^{-1} {\rm kpc}$}
\begin{document}

\title[Multiphase SPH] {Multiphase Smoothed-Particle Hydrodynamics} 
\author[B. W. Ritchie \& P.A. Thomas] {Benedict W. Ritchie\thanks{Email: B.W.Ritchie@sussex.ac.uk} 
and Peter A. Thomas\\
  Astronomy Centre, School of Chemistry, Physics and Environmental Science,
  University of Sussex, Falmer, Brighton BN1 9QJ}

\date{Accepted ---. Received ---; in original form ---}

\maketitle

\begin{abstract}
  
  We adapt the Smoothed-Particle Hydrodynamics (SPH) technique to allow a
  multiphase fluid in which SPH particles of widely differing density may be
  freely intermixed.  Applications include modelling of galaxy formation and
  cooling flows.

\end{abstract}

\begin{keywords}
  methods: numerical - hydrodynamical simulation - galaxies: formation -
  cooling flows
\end{keywords}

\section{Introduction}

Since its introduction more than two decades ago by Lucy (1977) and
Gingold \& Monaghan (1977), Smoothed-Particle Hydrodynamics (SPH) has
become one of the standard techniques for modelling astrophysical
fluid flow (e.g. Evrard 1988; Hernquist \& Katz 1989, hereafter HK89;
Thomas \& Couchman 1992; Steinmetz \& M\"{u}ller 1993; Couchman,
Thomas \& Pearce 1995, hereafter CTP95; Shapiro \etal 1996). SPH is
fully Lagrangian, with the particles themselves being the framework on
which the fluid equations are solved, and so there is no grid to
constrain the dynamic range or geometry of the system being modelled.
This is of particular importance for phenomena involving the growth of
gravitational fluctuations, such as cosmological structure formation
(Bertschinger 1998) and star formation (Bhattal \etal 1998), and in an
adaptive form (Wood 1981; Nelson \& Papaloizou 1994) the SPH algorithm
will follow the wide range of densities encountered without
difficulty. In contrast, computational demands severely restrict the
dynamic range of Eulerian finite-difference methods (e.g. the
Piecewise-Parabolic Method of Collela \& Woodward 1984), and to date
the best three-dimensional Eulerian simulations of galaxy formation
have a gas resolution of $\sim \! 300-500$ kpc (Blanton \etal 1999),
although adaptive mesh refinement (AMR) techniques promise to greatly
enhance the fixed-grid approach.  SPH can be easily integrated with a
range of $N$-body gravity solvers such as the tree algorithm used by
HK89 and the Adaptive Particle-Particle, Particle-Mesh (AP$^3$M)
algorithm of Couchman (1991).

However, SPH is not without its problems. The need for an artificial
viscosity means that SPH resolves shocks poorly in comparison with
finite-difference methods. Pairwise artificial viscosities (Monaghan
\& Gingold 1983) generally give the sharpest resolution of shocks, but
also introduce a large shear viscosity unless correction terms are
used (Balsara 1995). There is some concern that these terms further
degrade the shock capturing ability of SPH (Navarro \& Steinmetz 1997;
Thacker \etal 1998, hereafter T98).  While it is possible to add a
\textit{physical} viscosity to SPH and solve the Navier-Stokes
equations directly (e.g. Flebbe \etal 1994), the relatively simple SPH
interpolation method is quite sensitive to particle disorder and tends
to give large errors in the higher-order dissipative terms.  Boundary
conditions are also difficult to implement in SPH and do not sit
naturally with the method. Generally boundaries are either periodic or
situated far from the region of interest, and SPH is unsuitable for
simulations in which complex boundary conditions are of critical
importance.

Standard implementations of SPH also have a limited ability to resolve
steep density gradients, and a number of numerical problems can occur
when particles are close to, but not physically part of, a region of
higher density. These arise because the usual formulation of SPH
assumes that the density gradient across the smoothing kernel of each
particle is small. However, this is not true in many situations in
which SPH is commonly used. In simulations of galaxy formation, for
example, thermal instability causes cold, dense clumps to form within
halos of hot gas, and density contrasts of several orders of magnitude
can occur within the typical smoothing length of the halo gas. Current
implementations of SPH will overestimate the density of the halo gas,
leading to the gas cooling excessively and accreting on to the cold
clump (T98; Pearce \etal 1999).  Tittley, Couchman and Pearce (1999,
hereafter TCP99) have also shown that the drag exerted on these
protogalaxies by the intracluster medium can be seriously
overestimated.  These problems are thought to contribute to the
overmerging commonly seen in simulations of structure formation
including gas (Frenk \etal 1996).

We wish to extend the method to cope with these problems, along with
multiphase fluids, such as the intracluster medium and cooling flows,
which consist of an emulsion of discrete phases in which there is no
correlation between the density of neighbouring particles. We should
emphasise however that we are \textit{not} seeking to model
homogeneously mixed materials with different equations of state, such
as dust and aerosols (Monaghan 1997).  We make two changes to the
standard implementation of SPH. Firstly, particles no longer have to
try and maintain a fixed number of neighbours, as is required by the
algorithm of HK89, and instead the smoothing length is adjusted to
keep a density-weighted quantity constant (see
Section~\ref{sec:smooth2}).  This prevents the smoothing length of low
density particles from decreasing as they approach a high density
region. Secondly we make the assumption that pressure, not density, is
constant across the smoothing kernel. We then summate the local
pressure at each particle and calculate the local density from the
equation of state. This is a reasonable approach to take when the
cooling time of the gas is greater than the local dynamic time, where
a local pressure equilibrium can be expected even when the local
density gradient is steep. It will \textit{not} be true in the
presence of shocks (although neither is the assumption that density is
smoothly varying).  As we will see in sections~\ref{sec:shocktube}
and~\ref{sec:adiabatic}, however, there is no evidence that the shock
capturing ability of SPH is degraded.

The layout of the paper is as follows. In section~\ref{sec:method2} we detail
our new implementation of SPH. In Section~\ref{sec:test2} we present the
results of a number of tests of the new method, and compare the performance
with results produced using the SPH implementations of CTP95 and T98.
This is followed in Section~\ref{sec:disconc} by a discussion of our results.

\section{Methodology}
\label{sec:method2}

In SPH the fluid is represented by particles of known mass, $m_i$, and
specific energy, $\epsilon_i$.  Other properties must be inferred by averaging
over a smoothing sphere that typically extends to enclose a fixed number,
$N_{\rm SPH}$, of particles.  For a continuous distribution, the average value
of some quantity $A$ at the location of particle $i$ would be
\begin{equation}
\label{eq:adefcont}
\langle A_i\rangle = \int A(\vec{r})\,W(\vec{r}-\vec{r_i})\,{\rm d}V
\end{equation}
where $W$ is the smoothing kernel.  In SPH the integral is replaced by a sum:
\begin{equation}
\label{eq:adef}
\langle A_i\rangle = \sum_j {m_j\over\rho_j} A_j W_{ij}
\end{equation}
that extends over all particles, $j$, within the smoothing sphere.  $\rho_j$
is the density of particle $j$ and so $m_j/\rho_j$ is its volume. Here 
\begin{equation}
\label{eq:wdef}
W_{ij}={1\over h_{ij}^3}\,W(r_{ij}/h_{ij}),
\end{equation}
where $r_{ij}$ is the separation of particles $i$ and $j$,
$r_{ij}=|\vec{r}_i-\vec{r}_j|$, and $h_{ij}$ is the smoothing length.  There
are several possible forms for $W$.  Throughout this paper, we use the
standard form described in Thomas \& Couchman (1992). 
The choice of $h_{ij}$ determines the size of the region over which the
density is to be averaged.  Many authors take $h_{ij}$ to be a symmetric
function (for example the harmonic average) of $h_i$ and $h_j$ (see T99
and references therein).  We prefer to
set $h_{ij}=h_i$. This has the advantages that we know in advance exactly how
far we have to search and always have the same number of neighbours
within the smoothing region. The integral of the
smoothing kernel over all space is unity, which translates to the condition
\begin{equation}
\label{eq:consistency}
\sum_j {m_j\over\rho_j} W_{ij} = 1,
\end{equation}
although this will only be true in a statistical sense.

Equation~\ref{eq:adef} seems to
require the value of $\rho_j$, which is not known in advance.  However, it is
possible to circumvent this by formulating SPH such that $A$ is
always a multiple of density and a known quantity. Taking $A = \rho$ gives 
the standard SPH estimate for the density in the neighbourhood of a particle
\begin{equation}
\label{eq:dmean}
\langle\bar\rho_i\rangle = \sum_jm_jW_{ij}.
\end{equation}

Where expressions arise involving derivatives, we use the divergence theorem
to transfer the derivative onto the smoothing kernel as follows:
\begin{equation}
\label{eq:gradadef}
\langle\nabla A_i\rangle=\int \nabla\!A\,W\,\dd V
                        =-\int A\nabla W\,\dd V
\end{equation}
\begin{equation}
\label{eq:divadef}
\langle\nabla.\vec{A}_i\rangle=\int \nabla.\vec{A}\,W\,\dd V
                              = -\int \vec{A}.\nabla W\,\dd V
\end{equation}

\subsection{Evaluation of the density}
\label{sec:density}

In multiphase SPH, it is important to distinguish between the
density of a particle, $\rho_i$, and the mean density in the neighbourhood of
a particle, $\bar\rho_i$, defined by Equation~\ref{eq:dmean}. 
The pressure of the gas is expected to be a smooth quantity
because (excluding shocks) the sound-crossing time is
shorter than the flow time across the smoothing sphere. The
density of individual particles, $\rho_i$, can therefore be determined 
from an estimate of
the local pressure and the equation of state,
\begin{equation}
\label{eq:eos}
P = 2/3 \rho \epsilon
\end{equation}
The specific energy $\epsilon$ is related to
the gas temperature $T$ by $\epsilon = 3kT/2\mu m_H$, where 
$k$ is Boltzmann's constant, $m_H$ is 
the mass of the hydrogen atom and $\mu = 0.6$ is the relative molecular mass.
The SPH estimate of the pressure is 
\begin{equation}
\label{eq:pressure}
\langle P_i \rangle = 2/3 \sum_j m_j \epsilon_j W_{ij}
\end{equation}
and the density of a particle can therefore be written 
\begin{equation}
\label{eq:density}
\langle\rho_i\rangle = {3\langle P_i\rangle \over 2\epsilon_i}
={\sum_jm_j\epsilon_jW_{ij}\over\epsilon_i}.
\end{equation}
Variations in $\epsilon_i$ can cause large variations in density
between neighbouring particles, whereas Equation~\ref{eq:dmean}, being
an average, varies only slightly between neighbours. 
Note that a cold, high-density clump of particles will contribute a
lot of terms to Equation~\ref{eq:density}, but each of these will be
given a low weight due to the presence of the $\epsilon_j$ term. 

The consistency
condition, Equation~\ref{eq:consistency}, becomes
\begin{equation}
\sum_j{2m_j\epsilon_jW_{ij}\over3\langle P_j\rangle}=1
\end{equation}
which will be true if the pressure is slowly-varying, as we have assumed.

\subsection{The smoothing length}
\label{sec:smooth2}
The SPH smoothing length is usually defined in terms of the radius of a
sphere, centred on the particle in question, that encloses a fixed number of
neighbours.  One can either search for such a radius on each time-step, a
time-consuming process, or allow the number of neighbours to vary and
accept an estimate based on the number of neighbours found on the previous step:
\begin{equation}
\label{eq:hnew}
h_i\leftarrow h_i\,\left[\alpha + (1-\alpha)\left(N_{\rm SPH}\over
N_i\right)^{1\over3}\right],
\end{equation}
where $N_i$ is the actual number of neighbours and $N_{\rm SPH}$ is the
desired number (HK89).  $\alpha$ is a convergence parameter: for a uniform
distribution, choosing $\alpha=1$ returns the correct value of $h_i$ on the
next step.  Where large density contrasts are present, however, $\alpha$ must
be reduced to avoid overshooting and convergence is much slower. 
We use a convergence parameter $\alpha=0.4$ and $N_{\rm SPH} = 32$.  

For the extreme density contrasts we envisage, T98 have shown
that $N_i$ can oscillate between values which are alternately too low and too
high.  They solve this problem both by giving lower weight to particles at the
extreme edge of the smoothing sphere, and by the introduction of a more
sophisticated convergence algorithm.
In this paper, we suggest a simpler approach that uses a neighbour count
weighted as follows:
\begin{equation}
\label{eq:neigh}
N_i=\sum_j w_{ij}=\sum_j {2\bar{\rho}_i\over \bar{\rho}_i+\rho_j},
\end{equation}
where $\bar{\rho}_i$ is the mean density at particle $i$ and the sum
extends over all neighbours within the smoothing sphere. Note that:
\begin{enumerate}
\item For a uniform distribution in which $\bar{\rho}_i=\rho_j$, then 
$w_{ij}=1$ and $N_i$ is simply equal to the number of neighbours.
\item Particles do not notice neighbours with a much greater than
average density ($w_{ij}\sim0$ for
  $\bar{\rho}_i\ll \rho_j$).  This prevents the common instability whereby changing
  the radius of the smoothing sphere so as to include or exclude a
  high-density clump can dramatically change the number count.
\item Neighbours with a lower than average density count at most
  double ($w_{ij}\sim2$ for $\bar{\rho}_i
  \gg \rho_j$). While this could lead to an isolated cold particle in a hot 
  medium having fewer neighbours than is desirable, such a situation
  is unlikely
  to arise in practice and $w_{ij}$ could be limited to a maximum value
  of unity if desired. 
\end{enumerate}

\subsection{The equation of motion}
\label{sec:eom2}

In the absence of artificial viscosity, heating and radiative cooling, the
equation of motion for a parcel of gas is simply
\begin{equation}
\label{eq:eom2}
{{\rm d}\vec{v}\over{\rm d}t}=-\,{\nabla P\over\rho}.
\end{equation}
In the same spirit as above, we require an estimate of $\nabla P$ that does
not depend upon the number density of particles.  This is
\begin{equation}
\label{eq:gradp2}
\langle\nabla P_i\rangle={2\over3}\sum_jm_j\epsilon_j\nabla_iW_{ij}.
\end{equation}
It is common practice to use the following identity
\begin{equation}
\label{eq:gradpidentold}
{\nabla P\over\rho}=\nabla\left(P\over\rho\right)+{P\over\rho^2}\,\nabla\rho
\end{equation}
to symmetrise the pressure force.  This expression is not suitable here, both
because we are envisaging abrupt changes in density (so that $\nabla\rho$ is
undefined) and because the estimator for $P/\rho^2$ depends upon the density
of the particles. Equation~\ref{eq:gradpidentold} is obtained using the
general identity
\begin{equation}
\label{eq:gradpidentgen}
{\nabla P\over\rho}={P\over\rho^{\sigma}}
\nabla\left({1\over\rho^{1-\sigma}}\right)
+ {1\over\rho^{2-\sigma}} \nabla \left({P\over\rho^{\sigma-1}}\right)
\end{equation}
(Monaghan, 1992) and using $\sigma=1$ gives an alternative identity that at
first sight seems a little bizarre:
\begin{equation}
\label{eq:gradpidentnew}
{\nabla P\over\rho}={\nabla P\over\rho}+{P\over\rho}\,\nabla1.
\end{equation}
This leads to the estimator for the force on particle $i$,
\begin{equation}
\label{eq:force2temp}
\langle\vec{f}_i\rangle=\sum_j
{2\over3}m_im_j\left[
{\epsilon_j\nabla_iW_{ij}\over\langle\rho_i\rangle}
+{\epsilon_i\nabla_iW_{ji}\over\langle\rho_j\rangle}
\right],
\end{equation}
where we have chosen to use $W_{ji}$ in place of $W_{ij}$ in 
the second term on the right hand side in order to make the 
force symmetric. This gives
\begin{equation}
\label{eq:force2}
\langle\vec{f}_i\rangle=\sum_j\big(\vec{f}_{ij}-\vec{f}_{ji}\big),
\end{equation}
where
\begin{equation}
\label{eq:fij}
\vec{f}_{ij}={2\over3}m_im_j
{\epsilon_j\nabla_iW_{ij}\over\langle\rho_i\rangle}.
\end{equation}
The first set of terms in Equation~\ref{eq:force2} is evaluated when
calculating the SPH properties for particle $i$; the second set is accumulated
in stages when calculating the properties of the neighbours.
Note that the only change from the usual SPH formalism is that $\epsilon_i$
and $\epsilon_j$ have exchanged places in Equation~\ref{eq:force2temp}. 
However,
this change means that the force on particle $i$ is dependent only upon the
local pressure gradient and not upon the density of its neighbours.

\subsection{Conservation of energy}
\label{sec:energy2}

Conservation of energy is ensured by equating the heating to the $P\dd V$ work
done:
\begin{equation}
\label{eq:coe2}
{\dd\epsilon\over\dd t}=-{P\over\rho}\,\nabla.\vec{v}.
\end{equation}
By making use of the identity
\begin{equation}
\label{eq:divvident}
\nabla.\vec{v}=\nabla.\Delta\vec{v},
\end{equation}
where $\Delta\vec{v}=\vec{v}-\vec{v}_i$, this can be put into a variety of
forms.  For example,
\begin{equation}
{\dd\epsilon\over\dd t}=
-{\nabla.(P\,\Delta\vec{v})\over\rho}+{\Delta\vec{v}.\nabla P\over\rho}
\end{equation}
gives
\begin{equation}
\label{eq:dedt2one}
\left\langle{\dd\epsilon_i\over\dd t}\right\rangle=
{2\over3}\sum_j{m_j\epsilon_j\vec{v}_{ij}.\nabla_iW_{ij}
\over\langle\rho_i\rangle} = \frac{1}{m_i} \sum_j \vec{f}_{ij}.\vec{v}_{ij},
\end{equation}
where $\vec{v}_{ij}=\vec{v}_j-\vec{v}_i$ --the second term has vanished
because it contains $\vec{v}_{ii}$ which is identically zero.  Using
\begin{equation}
{\dd\epsilon\over\dd t}=
-{P\nabla.\Delta\vec{v}\over\rho},
\end{equation}
we obtain an alternative expression
\begin{equation}
\label{eq:dedt2two}
\left\langle{\dd\epsilon_i\over\dd t}\right\rangle=
{2\over3}\sum_j{m_j\epsilon_i\vec{v}_{ij}.\nabla_iW_{ij}
\over\langle\rho_j\rangle} = \frac{1}{m_i} \sum_j \vec{f}_{ji}.\vec{v}_{ji}.
\end{equation}
Combining Equations~\ref{eq:dedt2one} and \ref{eq:dedt2two}, we find that
\begin{equation}
\label{eq:dedt2}
m_i\left\langle{\dd\epsilon_i\over\dd t}\right\rangle=
{1\over2}\sum_j\big(\vec{v}_{ij}.\vec{f}_{ij}+\vec{v}_{ji}.\vec{f}_{ji}\big),
\end{equation}
where $\vec{f}_{ij}$ is the pairwise force given in Equation~\ref{eq:fij}.

Any of Equations~\ref{eq:dedt2one}, \ref{eq:dedt2two}, or \ref{eq:dedt2} (or
any similarly derived equation) may be used as an estimator of
$\dd\epsilon/\dd t$ (and hence, by rearrangement of Equation~\ref{eq:coe2},
for $\nabla.\vec{v}$). Indeed, if density is eliminated by substituting 
Equation~\ref{eq:eos} it is clear that the equations are identical, provided
that our assumption that $P_i$ and $P_j$ are approximately equal is valid. 
Overall energy conservation is ensured as all three expressions give
\begin{equation}
\frac{{\rm d}E}{{\rm d}t} = \sum_i \left( m_i \frac{{\rm d}\epsilon_i}
{{\rm d}t} + \vec{v}_i.\vec{f}_i \right) = 0.
\end{equation}
Under most conditions the performance of the
three equations is indistinguishable, and standard tests of SPH cannot tell
them apart. However, when large temperature variations are present
the double-sided form can lead to a large scatter in particle entropy. 
We use Equation~\ref{eq:dedt2one} here.  

\subsection{Artificial viscosity}

In the presence of shocks, we require a mechanism to convert relative motion
into heat.  In SPH, this is achieved via an artificial pressure, $Q$, that is
added to the usual one in regions of convergent flow.  This has the effect of
replacing $P$ by $P+Q$ in Equations~\ref{eq:eom2} and \ref{eq:coe2}.  In the
current method, we replace $\vec{f}_{ij}$ in Equations~\ref{eq:force2} and
\ref{eq:dedt2two} with $\vec{f}_{ij}+\vec{g}_{ij}$, where
\begin{equation}
\label{eq:artvis}
\vec{g}_{ij}=\vec{f}_{ij} {\cal M}_{ij}({\rm \beta}\,{\cal
M}_{ij}+{\rm \alpha})
\end{equation}
(c.f. Monaghan \& Gingold 1983). Here the pairwise Mach number
\begin{equation}
\label{eq:mach}
{\cal M}_{ij}=\cases{
0,&     $\vec{r}_{ij}.\vec{v}_{ij}>0$\cr
{h_i|\vec{r}_{ij}.\vec{v}_{ij}|\over c_{ij}\,(r_{ij}^2+0.01h_i^2)},&
$\vec{r}_{ij}.\vec{v}_{ij}<0$
}
\end{equation}
the particle separation, $r_{ij}=|\vec{r}_{ij}|=|\vec{r}_j-\vec{r}_i|$, and
the sound speed 
\begin{equation}
\label{eq:sound}
c_{ij}= \frac{c_i + c_j}{2}
\end{equation}
where
\begin{equation}
c=\surd(10\epsilon/9).
\end{equation}
We adopt the values $\alpha=1$, $\beta=2$. 
The use of $c_{ij}$ in Equation~\ref{eq:mach} rather than
$c_i$ helps to limit the degree to which cold particles shock
against dense clumps. 

In addition, a shear-correcting term (Balsara 1995) can
be applied to limit the damping 
of shear flows. Following Steinmetz (1996) we use
\begin{equation}
{\cal M}_{ij} \to \tilde{\cal M}_{ij} = {\cal M}_{ij} k_i
\end{equation}
where
\begin{equation}
k_i = \frac{| \langle \nabla . \vec{v} \rangle_i |}{| \langle \nabla .
\vec{v} \rangle_i | + | \langle \nabla \times \vec{v} \rangle_i | 
+ 0.0001 c_i / h_i}.
\end{equation}
For compressional flows $k$ = 1 and ${\cal M}_i$ is unchanged, while
in shearing flows $k \to 0$ and the artificial viscosity vanishes. 
Only the cosmological galaxy formation test described in 
Section~\ref{sec:cosmic} uses this correction term. 

\subsection{Cooling}
\label{sec:cool2}

In the presence of artificial viscosity and cooling, the Equation of Energy
Conservation becomes
\begin{equation}
\label{eq:coole}
{\dd\epsilon\over\dd t}=-{P+Q\over\rho}\,\nabla.\vec{v}-\xi,
\end{equation}
where $\xi$ is the emissivity (the emission rate per unit volume).  
The cooling function is interpolated from Sutherland \& Dopita (1993). To
minimise problems of cooling during shock-heating (see Hutchings \& Thomas
2000),
we allow the gas to cool only at the end of a time-step after the artificial
viscosity term has been applied.  The cooling is assumed to occur at constant
density (the time-step ensures that this condition is approximately 
satisfied), as described in Thomas \& Couchman (1992). 

\section{TESTS}
\label{sec:test2}

The tests discussed in the following sections were carried out using
Hydra\footnote{This code is in the public domain and can be downloaded from
http://hydra.mcmaster.ca/.} (Couchman, Thomas and Pearce, 1995), an 
AP$^3$M+SPH code modified to use the SPH method detailed in this paper. The
simulations were performed on single-processor Sun and Intel workstations. \\

\subsection{Shock tube}
\label{sec:shocktube}
\begin{figure*}
\begin{minipage}{17.6cm}
 \begin{center}
   \psfig{file=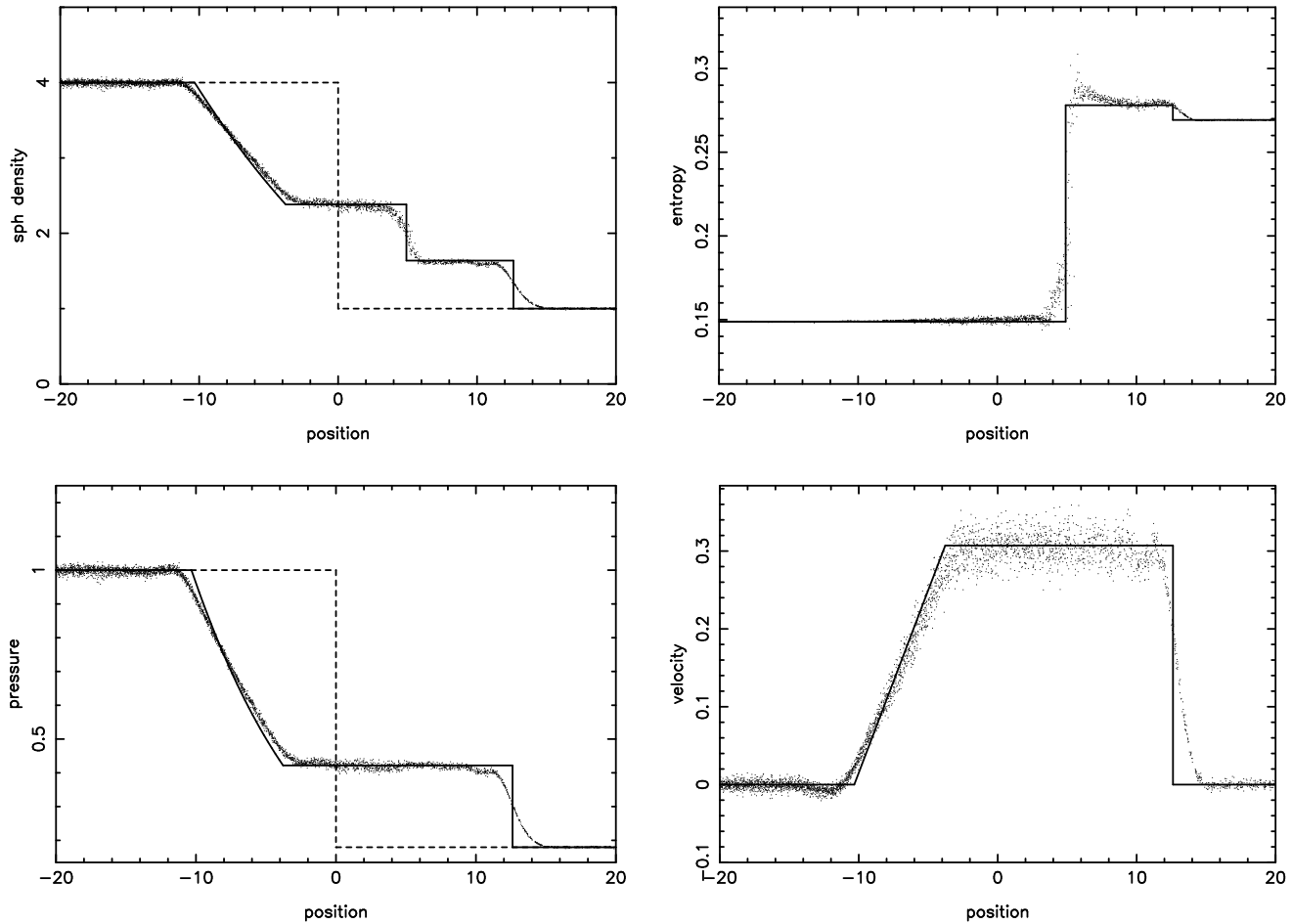,width=17.4cm}
   \caption{Results from the Sod shock test after 30 time-steps. The variation
     in density, pressure, velocity and the entropic function $A(s) =
     \varepsilon / \rho^{(\gamma-1)}$ across the shock are shown. Analytic
     solutions are shown as solid lines, and the initial pressure and density
     profiles defined in equation~\ref{eq:sodinitial} are shown as dashed
     lines. The smoothing length in the vicinity of the shock is $\sim 0.9$ 
        in code units.}
  \label{fig:sodshock1}
 \end{center}
\end{minipage}
\end{figure*}

A standard test of gas dynamical codes is the Sod shock (Sod 1978),
which has been applied to SPH by many authors (e.g. Monaghan \&
Gingold 1983; Rasio \& Shapiro 1991; T98).  Analytic results are given
by Hawley, Smarr \& Wilson (1984) and Rasio \& Shapiro (1991). This
test is often carried out in one dimension, but this does not properly
test particle interpenetration; we perform a three dimensional test,
with the normal cubical simulation volume altered to match the
geometry of a shock tube. Dimensions are $6\times6\times120$ in code
units and all boundaries are periodic.

In a Sod shock two regions of gas with
different densities are brought into contact, resulting in a shock wave
propagating into the low-density gas and a rarefaction wave propagating into
the high density gas.  Between these two regions is a contact discontinuity,
where the pressure is constant but the density jumps. Following T98 we 
use the initial conditions
\begin{equation}
\begin{array}{lllr}
\rho_l = 4 & P_l = 1      & v_l = 0. & {\rm for} \; x < 0 \\
\rho_r = 1 & P_r = 0.1795 & v_r = 0. & {\rm for} \; x \geq 0 
\end{array} 
\label{eq:sodinitial}
\end{equation}
giving a shock Mach number ${\cal M}\!\! \sim \!\! 1.4$. Both regions are
allowed to evolve at constant temperature before being brought into contact,
to allow the gas to relax to a physically realistic initial state.  A total of
7343 equal mass particles are used.

\begin{figure}
 \begin{center}
   \psfig{file=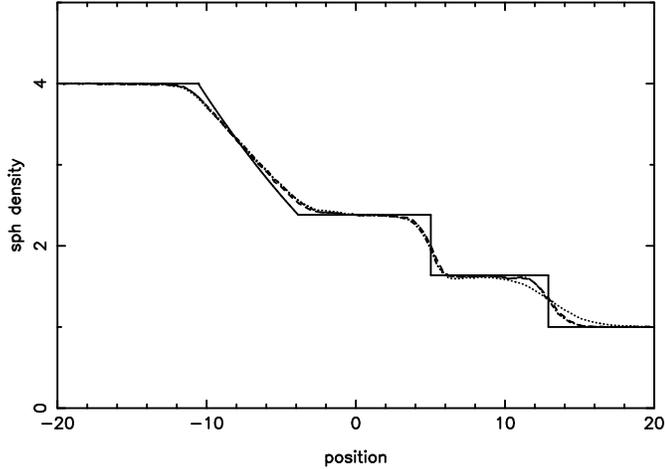,angle=270,width=8.7cm}
  \caption{A comparison between the results from this code (dashed line) and
    those of CTP95 (dotted line) and T98 (dotted-dashed line). The shock front
    is at $x \approx 13$. The broader shock produced by the CTP95 code
   is a result of the poor shock capturing of the artificial viscosity
   used.}
  \label{fig:sodcompare}
 \end{center}
\end{figure}

Thirty timesteps are required to reach the state shown in
Figure~\ref{fig:sodshock1}, by which time the shock front has moved
around 13 code units to the right.  Both the shock and the contact
discontinuity are broadened over a range $\Delta x \approx 3h$. The
results are in good agreement with the analytic solutions, although in
common with other implementations of SPH the gradient of the
rarefaction wave is too shallow. Flow is reasonably smooth, and there
is no post-shock ringing.  Figure~\ref{fig:sodcompare} compares the
results from our code with those of T98 and CTP95.  There is very
little difference between the three implementations, although the
CTP95 implementation produces a broader shock than the other two
codes.  This is due to the choice of artificial viscosity -- CTP95
uses an artificial viscosity based on the divergence of the local
velocity field, which is shown in T98 to be worse at shock capturing
than the pairwise artificial viscosity used in T98 and our code. All
three implementations model the rarefaction wave similarly, as viscous
terms do not apply in expanding regions.

With ${\cal M}\!\!\sim\!\! 1.4$, this test represents a fairly weak
shock. A strong adiabatic shock presents a more demanding test, as the
pressure jump across the shock, and hence the Mach number, are
infinite. In this case the jump conditions are
\begin{equation}
\begin{array}{lllr}
\rho_l = 1 & P_l = 0     & v_l = 1.333 & {\rm for} \; x < 0 \\
\rho_r = 4 & P_r = 1.333 & v_r = 0.333 & {\rm for} \; x \geq 0 
\end{array} 
\label{eq:adinitial}
\end{equation}
although in our code we require particles to have a small minimum
temperature to prevent numerical divergences in Equation~\ref{eq:density}, 
leading to $P_l$ being slightly greater than zero and the Mach number 
remaining large but finite.  
\begin{figure}
 \begin{center}
   \psfig{file=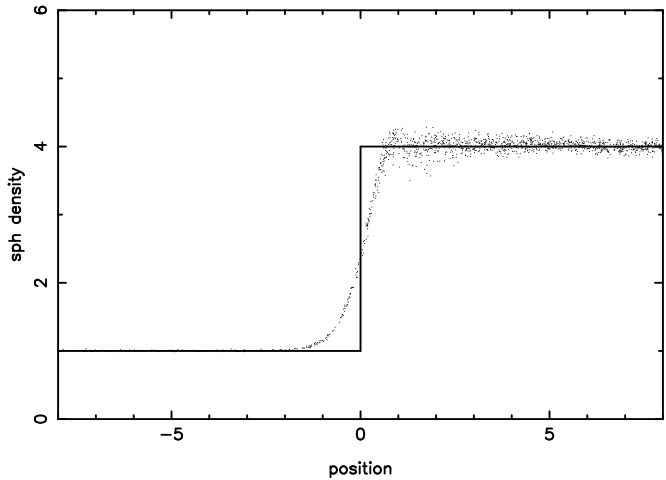,angle=270,width=8.7cm}
  \caption{The density jump across a strong adiabatic shock. Results
   from this code are shown. The smoothing length is $\sim 0.8$ on the
   low-density side of the shock, and $\sim 0.5$ on the high density
   side, both in code units.}
  \label{fig:addn}
 \end{center}
\end{figure}
The density profile across the shock is shown in Figure~\ref{fig:addn}. 
The new code is clearly capable of handling such shocks, and although 
some post-shock oscillation is visible it is not noticeably different from the
results obtained from the other codes. As the performance of our code is 
so close to that of T98 there is no evidence from shock-tube
experiments that our assumption that 
pressure is smoothly varying has degraded the shock-capturing ability of 
the code.

\subsection{Adiabatic collapse}
\label{sec:adiabatic}

One of the primary requirements for any hydrodynamics code that includes
gravity is the ability to correctly follow the shock-heating of cold gas during
gravitational collapse. A common test problem for SPH codes is the collapse
of an initially isothermal sphere of gas (Evrard 1988; HK89; Steinmetz \& 
M\"{u}ller 1993; T98), with an initial density profile
\begin{equation}
\label{eq:evrard}
\rho(r) = \frac{M(R)}{2\pi R^2}\frac{1}{r}.
\end{equation}
To create this profile we translate particles radially from a
uniform grid, which gives a lower sampling error than a
random distribution of particles. Initial particle temperatures are
set to the code minimum. A total of 8184 particles are used,
with the gravitational softening set as 0.02R. Following Evrard 
(1988) results are
presented in normalised units, with density, internal energy,
velocity, pressure and time normalised by $3M/4\pi R^3$, $GM/R$,
$(GM/R)^{1/2}$, $\rho u$ and $(R^3/GM)^{1/2}$ respectively.

\begin{figure}
 \begin{center}
   \psfig{file=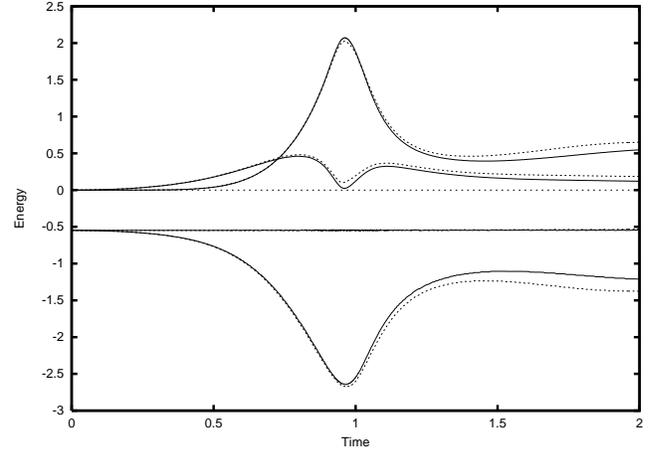,angle=270,width=8.7cm}
  \caption{Time evolution of the energies in the adiabatic collapse test. 
        Results from the multiphase (solid line) and standard 
        (dashed line) methods are shown, 
        and the zero point is marked with a dotted 
        line. The curves are, from top to bottom, the
        thermal, kinetic, total and potential energies. 
        Normalised units are used.}
  \label{fig:evenergy}
 \end{center}
\end{figure}
The gas in the sphere initially has negligible thermal energy, and
collapses due to the lack of pressure support. During the collapse a
central bounce occurs, causing a shock wave to propagate outwards
through the gas, and the sphere should eventually reach virial
equilibrium, with the ratio of thermal and gravitational energy
approaching a value of $-0.5$.  Figure~\ref{fig:evenergy} compares the
evolution of the thermal, kinetic, total and gravitational potential
energies during the collapse for the multiphase and standard methods.
The performance of the two codes is very similar throughout the bounce
and subsequent expansion, with only relatively minor differences
apparent after time $t\sim 1$.

\begin{figure*}
\begin{minipage}{17.6cm}
 \begin{center}
   \psfig{file=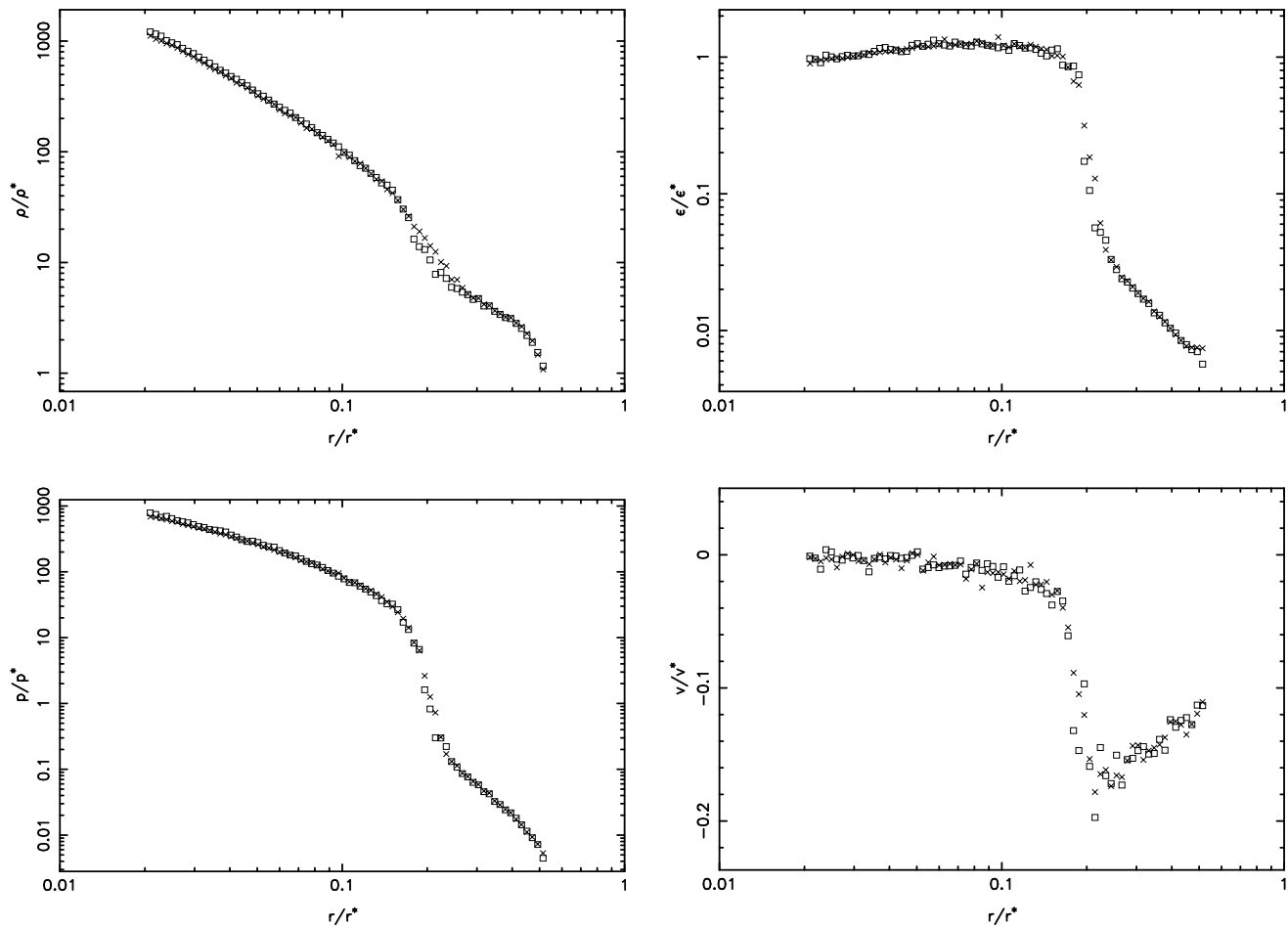,width=17.4cm}
   \caption{Radial density, pressure, internal energy 
   and velocity profiles for
  the adiabatic collapse test at time $t = 0.8$. The shock is located at
   $r/r^* \approx 0.2$. Results from this code are shown with '$\times$'
   symbols and results from a standard SPH code are shown with '$\Box$'
   symbols. Normalised units are used.} 
  \label{fig:evcollapse}
 \end{center}
\end{minipage}
\end{figure*}
Profiles of the system at time $t=0.8$ are shown in
Figure~\ref{fig:evcollapse}, at which time the shock is located at
$r/r^* \approx 0.2$. The temperature jump across the shock is well
modelled by both codes, and the post-shock conditions are very
similar. However, the density profile across the shock produced by the
multiphase code is noticeably shallower. This is a result of the
overestimation of the local pressure due to the steep pressure
gradient in the shock, and is analogous to the smoothing of the
density profile in the presence of steep density gradients that
affects the standard implementation of SPH.  This error is unimportant
in the force calculations, which are dominated by the artificial
viscosity during shocks, but may be significant when radiative cooling
is implemented, as it leads to to excess cooling in the shock.

In such circumstances it may be preferable to use the standard SPH
estimate of the density $\bar{\rho}_i$ for calculating the emissivity
in regions where the local pressure gradient is steep. This switch can
be implemented easily using either a pairwise condition based on the
pressure of particles $i$ and $j$ or an SPH estimation of the local
pressure gradient. Both methods effectively restrict the use of
$\bar{\rho}_i$ to the vicinity of the shock.  When this approach is
used, the density profile across the shock in
Figure~\ref{fig:evcollapse} closely resembles that of the standard
code, as would be expected, with the other profiles remaining
unchanged.

\subsection{Density estimation in a two-phase medium}
\label{sec:hupdate}

In the standard formulation of SPH arbitrarily steep density gradients
are smoothed over a distance representative of the local value of $h$.
This is a cause for concern in simulations of cosmological structure
formation, where cold dense clumps of gas -- galaxies -- form within
diffuse halos of hotter gas and the density of halo gas can be
overestimated by an order of magnitude or more in the steep density
gradients around the largest galaxies.

The smoothing of the density is a result of two effects. Firstly, 
when Equation~\ref{eq:dmean} is used, 
$\bar\rho \propto h^{-3}$ 
for a fixed number of neighbours, and the estimated 
density becomes closely tied to the choice of smoothing length. 
Secondly, most adaptive forms of SPH update the smoothing length 
each timestep to try and keep the number of neighbouring particles
approximately constant (e.g. the method of HK89). 
Particles act as tracers of the underlying 
mass distribution, and a high-density region will
therefore contain more particles than a region of lower density. 
When the HK89 algorithm is used, a particle
close to a dense clump of gas will need to search little further 
than the edge of the dense region to find its required number 
of neighbours. If Equation~\ref{eq:dmean} and the HK89 algorithm 
are used together then the estimate of the density of the particle 
becomes strongly dependent on its separation from the clump. 

Simply estimating the density using Equation~\ref{eq:density} is 
not enough to cure this problem. While the high density particles will 
be at low temperature in a region of pressure equilibrium, and will 
therefore contribute little to the estimate of the pressure, the 
size of the smoothing sphere, and hence weight given to 
neighbouring particles by the smoothing kernel $W_{ij}$, still 
depends on the distance to the clump. It is therefore necessary 
to weight the neighbour count using 
Equation~\ref{eq:neigh}, so that a low-density particle close 
to a dense clump will assign a very low weight to particles in
the clump and will search for neighbours as if the clump was not
present. 

\begin{figure*}
\begin{minipage}{17.6cm}
 \begin{center}
   \psfig{file=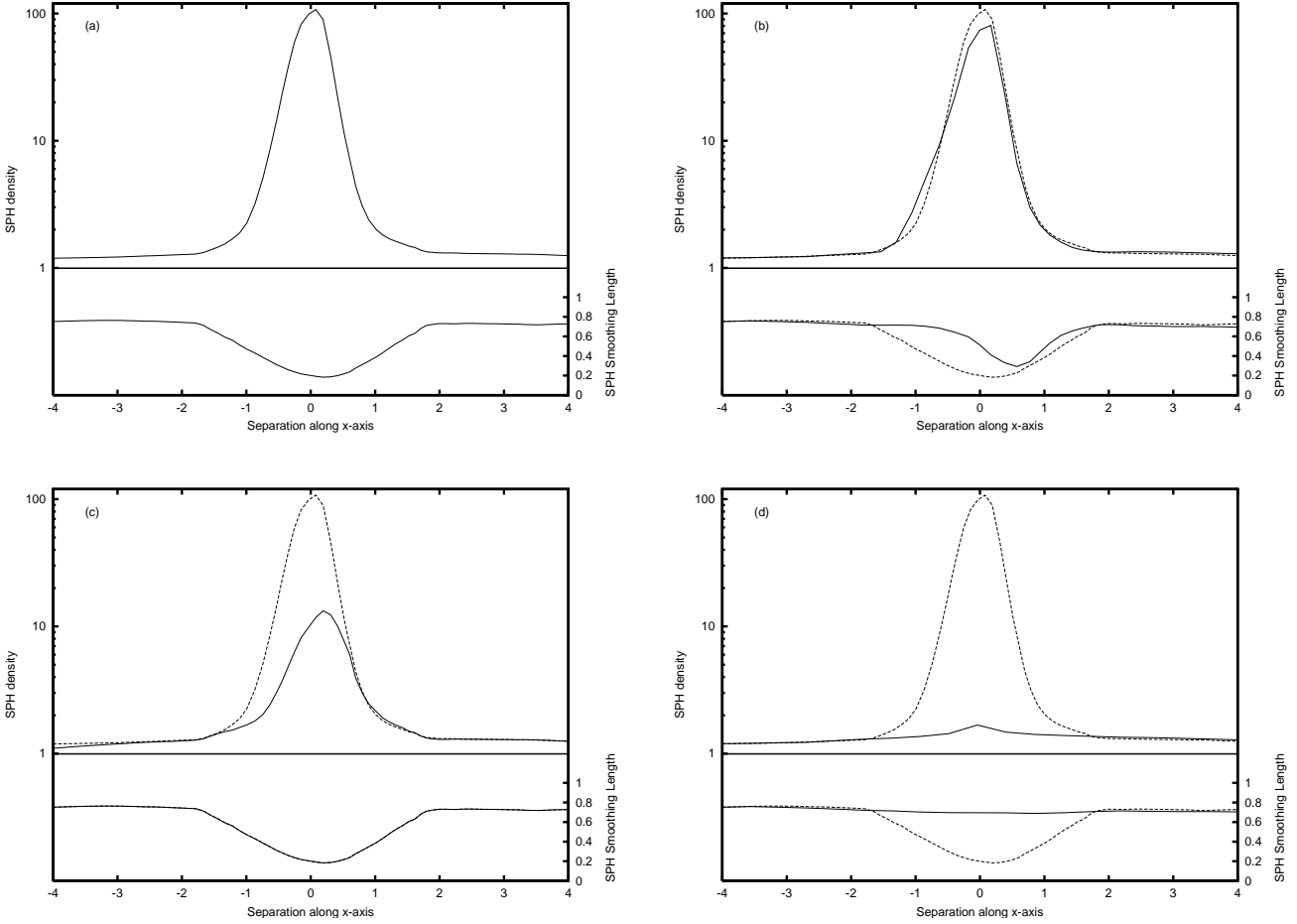,width=17.4cm}
   \caption{Change in the SPH estimate of the density $\rho$ and 
     the smoothing length $h$ as a particle in a stream of hot gas
     moves past a cold, dense clump, grazing the surface of the clump
     at closest approach. The central density of the clump is 200
     times that of the surrounding gas, with the temperature set so
     that the clump is in pressure equilibrium with its surroundings.
     Panel (a) shows results from the standard SPH approach, with
     density estimated using Equation~\ref{eq:dmean} and the smoothing
     length updated using the method of HK89. These results are
     plotted using dashed lines in subsequent panels. Panel (b) shows
     the effect of our new $h$-update algorithm when combined with the
     standard estimation of the density. Panel (c) shows the
     combination of the new estimate of the density given by
     Equation~\ref{eq:density} and the HK89 $h$-update method.
     Finally, panel (d) shows results from combining the new estimate
     of the density with the new $h$-update algorithm.}
  \label{fig:htest}
 \end{center}
\end{minipage}
\end{figure*}

The benefits of our approach can be seen in Figure~\ref{fig:htest}. Here
we plot the change in smoothing length and the SPH estimate of the
density as a hot gas particle moves past a cold, dense clump, 
grazing the surface of the clump at closest approach. The
clump contains 420 particles, and 
has a central density 200 times that of the surrounding gas,
with temperatures set so that the two phases are in pressure
equilibrium. For the purposes of this test all inter-particle forces 
have been turned off, so that the particles move at constant velocity. 
Initially, the density of the hot particle is $\sim 1.15$ and 
the smoothing length $h \sim 0.75$, both in code units. 

Panel (a) shows the results from the  standard implementation of SPH,
in which the density is estimated using Equation~\ref{eq:dmean} and
the smoothing length is updated using the algorithm of HK89. 
Once the smoothing kernel of the hot particle overlaps the cold
clump it finds many more than the desired number of neighbours
and the HK89 algorithm starts to decrease the smoothing length in response,
with $h$ reaching a minimum of $\sim 0.19$ shortly after closest
approach (this delay is due to the convergence parameter $\alpha$ 
in Equation~\ref{eq:hnew} limiting the rate at which $h$ can
change). There is a corresponding increase in the estimate of the
density, which peaks slightly earlier at $\rho \sim107$, an overestimate of 
some two orders of magnitude. These
values of $h$ and $\rho$ are typical of cold particles on the surface
of the clump, indicating that despite the large difference in
temperature between the hot and cold particles the standard
implementation of SPH treats them identically.

Panel (b) shows the result of changing to the $h$-update method
described in Section~\ref{sec:hupdate}, with the density still estimated
using Equation~\ref{eq:dmean}. The peak density has only dropped slightly
to $\rho \sim 80$, while the smoothing length remains fairly constant
until shortly before the closest approach, when the density of the hot
gas particle has increased to a point where particles in the clump become
significantly weighted, leading to a decrease smoothing length. 

Panel (c) shows the combination of the new estimate of the density,
given by Equation~\ref{eq:density}, and the HK89 $h$-update
algorithm. The peak density is reduced to $\rho \sim 10$, which occurs
when the smoothing length reaches it's minimum and the cold particles 
in the clump are weighted most strongly.

Finally, panel (d) shows the results when the new estimate of the
density is combined with the new smoothing length update algorithm. 
The smoothing length remains virtually constant
throughout the transit, with $h$ decreasing to $\sim 0.72$
shortly after closest approach and $\rho$ only increasing from $1.15$ to
$1.61$. 

The new $h$-update algorithm does imply some additional computational
expense when strong density gradients are present. In the example
presented here the hot particle has in excess of 450 neighbours when
the smoothing sphere overlaps the whole of the cold clump, far more
than the 32 required by the HK89 algorithm. The overall computational
expense will not be nearly as severe as this might suggest, however,
as this idealised test follows a single particle next to a dense
clump, where the weighting of neighbouring particles is the most
extreme. For a distribution of particles, such as the cosmological
structure formation discussed in Section~\ref{sec:cosmic}, the
overhead is typically on the order of $\sim 10\%$ per timestep.

\subsection{Force estimation in a two-phase medium}
\label{sec:twophase}

\begin{figure}
 \begin{center}
   \psfig{file=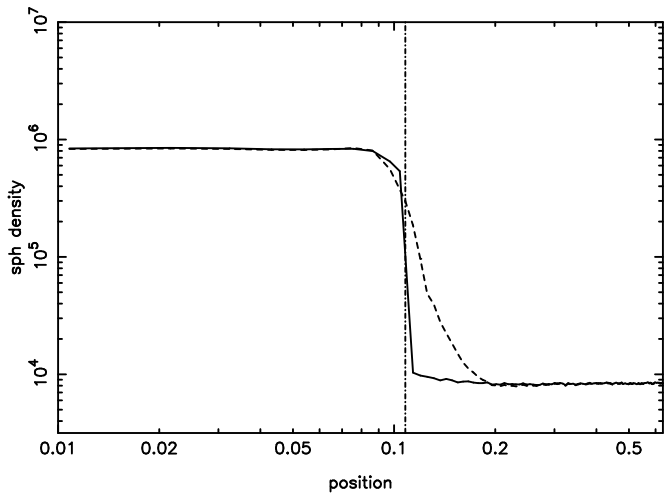,angle=270,width=8.7cm}
   \caption{SPH estimation of the density across a density contrast of
    $\rho_l / \rho_r = 100$, which occurs at $r=0.108$. Results from our new
    algorithm are shown as a solid line, while the results from the algorithm
    of T98 are shown as a dashed line. All units are code units.}
  \label{fig:dncomp}
 \end{center}
\end{figure}

Figure~\ref{fig:dncomp} shows the SPH estimate of the density profile across 
the boundary between two regions of different density, and demonstrates 
the sharpness with which density contrasts can be resolved by our method. 
To create the high density region we first evolve a cubical volume of
gas at constant temperature, to ensure a relaxed particle distribution. 
We then extract a spherical region and compress it radially to achieve 
the desired density. This is then inserted back into the cubical simulation
volume and the particles are allowed to move at constant radius from the 
centre of the box for a few time-steps to ensure a fully stable initial
state. Here the sphere has a radius $r=0.108$ in code units, and is
100 times more dense than the surrounding gas, with the temperature 
again set so that the two regions are in pressure equilibrium.
Away from the boundary both methods correctly estimate the density, as we
would expect. However, as discussed in the previous section, the standard 
implementation of SPH clearly overestimates the density close to the dense
region, whereas the new algorithm gives a density contrast that is much
sharper. 

In addition to the problems in estimating the density, 
unphysical forces can also occur in the
presence of steep density gradients. In implementations of SPH which try to
keep the number of neighbours constant, a particle near a density gradient
will find many, if not all, of its neighbours in the region of higher density.
The pressure gradient at the particle will therefore be highly asymmetric,
leading, from Equation~\ref{eq:eom2}, to a force that acts to push the
particle away from the high density region. The strength of this force will
depend on the magnitude of the density contrast, saturating once the particle
finds all its neighbours in the high density region, and a dense clump of gas
will therefore rapidly empty the surrounding region of particles.

\begin{figure}
 \begin{center}
   \psfig{file=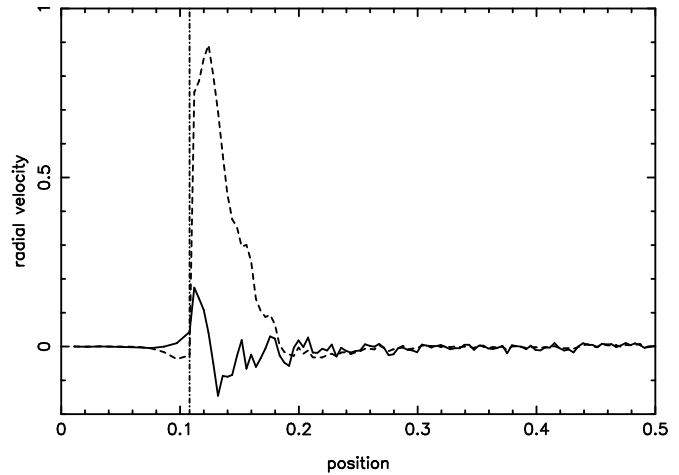,angle=270,width=8.7cm}
  \caption{Average radial velocity after 10 time-steps across
    a density contrast $\rho_h / \rho_l = 100$, which occurs at $r=0.108$.
    Results from our code are shown as a solid line, and results from the code
    of T98 are shown as a dashed line}
  \label{fig:veldiff}
 \end{center}
\end{figure}

We can see the effect of this asymmetric pressure gradient in 
Figure~\ref{fig:veldiff}, where we plot the radial velocities after
10 time-steps of particles in or near the dense clump. 
\begin{figure}
 \begin{center}
   \psfig{file=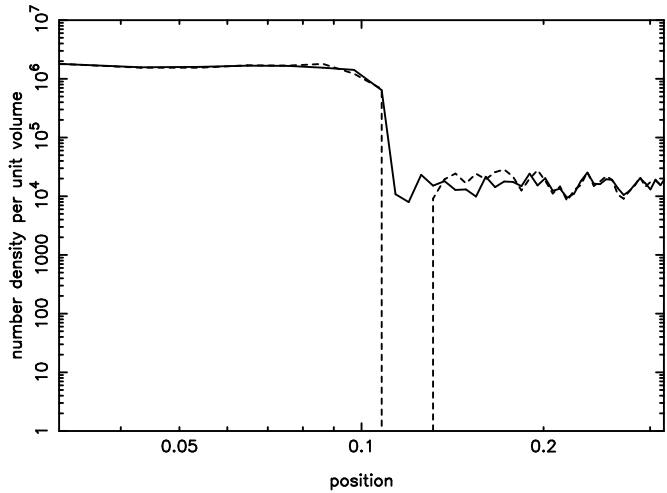,angle=270,width=8.7cm}
  \caption{Number density of particles per unit volume after 25 time-steps 
    across a density contrast $\rho_h / \rho_l = 100$, which occurs at
    $r=0.108$.  Results from our code are shown as a solid line, and
    results from the code of T98 are shown as a dashed line. The line
    marking the boundary between the two regions has been omitted for
    clarity. The empty region around the dense central clump can be
    clearly seen.}
  \label{fig:numdens}
 \end{center}
\end{figure}
The scatter in velocity due to thermal motion of the hot gas is around
$\pm 0.05$ in both cases.  The standard SPH code produces a large
outward velocity in the hot gas, which evacuates a space between the
two phases within a few tens of time-steps. This effect can be clearly
seen in Figure~\ref{fig:numdens}, where the number density of
particles drops to zero between $0.108 \leq r \leq \!  0.13$. While
our method also produces excess velocities at the boundary, implying
that the spurious pressure gradient has still not been completely
eliminated, the magnitude of the outward velocity has been greatly
reduced and the effect is far less systematic.

\subsection{Drag}
\label{sec:drag}

Drag resulting from gas dynamical forces is an important factor in
simulations of cosmological structure formation, as incorrectly
estimating the drag can bias both the distribution of matter in
clusters and the size and number of objects formed. Frenk \etal (1996)
have suggested that excessive drag can worsen the overmerging problem
seen in N-body simulations, and TCP99 have shown that standard
implementations of SPH can significantly overestimate the drag on a
cold clump of gas moving through hot gas representative of the
intracluster medium. This excess drag is caused by accretion of gas
from the ICM on to a shell around the clump, where it is held by
forces arising from the miscalculation of the pressure gradient around
the clump (a discussion of this effect can be found in TCP99).  This
accretion leads to an increase in the effective radius of the clump,
and hence the drag. This drag is most severe when the clump is moving
subsonically, and at higher velocities the accretion of gas decreases,
largely vanishing when the Mach number ${\cal M} \ge 2$.

\begin{figure}
 \begin{center}
   \psfig{file=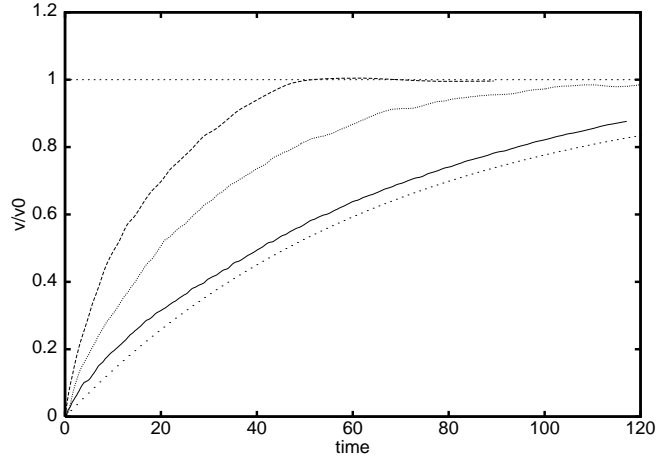,angle=270,width=8.7cm}
  \caption{Rate of acceleration of a dense clump in a flow with Mach 
    number 0.5. Results from our code are shown as a solid line,
    results from T98 are shown as a dashed line, results from CTP95
    are shown as a densely dotted line and the prediction of
    Equation~\ref{eq:drag} are shown as a sparsely dotted line.  All
    codes produce an excess drag.}
  \label{fig:drag}
 \end{center}
\end{figure}

In order to examine the drag introduced by our implementation of SPH,
we consider the case of a clump of cold gas initially at rest in a
stream of fast moving, diffuse gas (this is identical to the method of
TCP99, except we work in the rest-frame of the cold clump). Assuming
collisions are inelastic, the clump will be accelerated to the flow
velocity $v_0$ at a rate
\begin{equation}
        v = v_{0} \left [ \frac{e^{k t}-1}{e^{k t}+1} \right ]
\label{eq:drag}
\end{equation}
where $k$ is a constant
\begin{equation}
        k = \frac{2 \pi r^2 \rho_0 v_0}{M} 
\end{equation}
in which $\rho_0$ is the density of the diffuse gas, $v_0$ the flow
velocity, r is the radius of the clump and M is the mass of the clump
(all in code units).  For our tests $M=50$, $r = 0.5$, the velocity
and density of the flow are set to unity and the Mach number of the
flow is determined by setting the temperature of the gas as required.
We use a Mach number ${\cal M} \! \sim \!  0.5$ here, which is well
inside the regime in which TCP99 show that drag becomes excessive.
Figure~\ref{fig:drag} compares the results from our new code with
those obtained the codes of CTP95 and T98 (the code used by TCP99 is
essentially the same as that used in T98).
The results of our code are clearly an improvement, being close to the
prediction of Equation~\ref{eq:drag}. This is due to the elimination
of the accretion that occurs in the other codes, as the asymmetric
pressure gradients causing accretion are greatly reduced by our
method.  The difference between CTP95 and T98 is mainly a result of
the kernel smoothing used in T98, which is shown in TCP99 to further
increase the effective cross section of a dense clump. At higher Mach
numbers the differences between the codes are less pronounced as the
accretion decreases once ${\cal M} > 1$, although T98 continues to
give the largest drag of the three codes.

\subsection{Multiphase cooling flows}
\label{sec:cflow}

Clusters of galaxies contain large quantities of hot, X-ray emitting
gas, often concentrated around the most massive galaxy in the cluster.
In $\sim \! 70\% - 90\%$ of cases, gas in the centre of the cluster
has a radiative cooling time less than the Hubble time, $H_0^{-1}$
(Edge, Stewart \& Fabian 1992; White, Jones \& Forman 1997), and as
this gas cools surrounding gas will move inwards to maintain pressure
support, initiating a large-scale motion known as a cooling flow (see
Fabian 1994 for a review).  The mass deposition rate can be as high as
$10^3 M_{\odot} {\rm yr}^{-1}$ (Fabian \etal 1985), and is often
observed to vary with radius roughly as $\dot{M}(<r) \propto r$ within
the radius $r_{{\rm cool}}$ in which the cooling time is less than a
Hubble time (e.g. Thomas, Fabian and Nulsen 1987, hereafter TFN87).
This is generally taken as evidence for a multiphase flow, in which
thermal instability is causing the denser gas to cool out of the flow
at larger radii.

Modelling a multiphase flow is impossible with the usual
implementation of SPH as density is a locally averaged quantity,
making the flow inherently single phase. In contrast, our method
allows a wide range of densities, provided that a local pressure
equilibrium exists. This is a reasonable assumption for particles with
temperatures above $10^6$K, which will have cooling times much greater
than the local sound crossing time (Nulsen, 1988).  Particles below
$10^6$K will cool to $10^4$K within a few time-steps, at which point
they are removed from the flow.

To test the ability of our code to model a fully multiphase
environment, we examine a simple constant-pressure,
spherically-symmetric cooling flow.  The distribution of phases in the
flow is described by the fractional volume distribution $f(\rho,r)$
introduced by Nulsen (1986, hereafter N86), where $f {\rm d}\rho$ is
the fractional volume occupied by phases in the density range $\rho$
to $\rho + {\rm d}\rho$.  N86 considered many analytic forms for $f$.
Here, we take
\begin{equation}
f(\rho,r) = \frac{(3-\alpha)}{\rho_0}\left(\frac{\rho}{\rho_0}\right)^{-(4-\alpha)} \hspace{1cm}
\rho > \rho_0
\label{eq:cffuse}
\end{equation}
with $\rho_0$ being a minimum density and $\alpha$ the temperature 
dependence of the cooling 
function $\Lambda \propto  T^{\alpha}$. For the purposes of this test
we replace the Sutherland \& Dopita (1993) cooling function with a pure
power law in which  $\alpha=0.5$. 
Then Equation~\ref{eq:cffuse} can be integrated (N86) and gives 
a mass deposition profile
\begin{equation}
\dot{M} \propto r^{\eta}
\end{equation}
where
\begin{equation}
\eta = \frac{3(2-\alpha)}{(5-2\alpha)} = \frac{9}{8}.
\label{eq:massdepo}
\end{equation}

Particles are placed randomly within a cubical simulation of volume
$(200{\rm kpc})^3$, and are then allowed to evolve at constant
temperature until spurious fluctuations arising from the initial
particle distribution have died away. Particles are then ordered in
terms of their distance from the centre of the box, and translated
radially so as to match the mass profile given by
\begin{equation}
M \propto \rho_0(r) r^3
\end{equation}
where the density profile
\begin{equation}
\rho_0(r) \propto r^{-\frac{3}{(5-2\alpha)}} \propto r^{-3/4}
\label{eq:cfdens}
\end{equation}
is derived in TFN87. Particles translated outside the bounds 
of the box are discarded. 
Particle densities are drawn at random from the 
volume fraction distribution given by Equation~\ref{eq:cffuse}, 
with particle temperatures set so as to 
maintain constant pressure. Here, the outer temperature 
is $T \sim 5\times10^7K$, 
the inner temperature
$T \sim 10^6 K$, the central density is $0.1 \rm{cm}^{-3}$ and the average 
outer density is $\sim 5 \times 10^{-3} \rm{cm}^{-3}$. A total of 
20252 particles
are used.

\begin{figure}
 \begin{center}
   \psfig{file=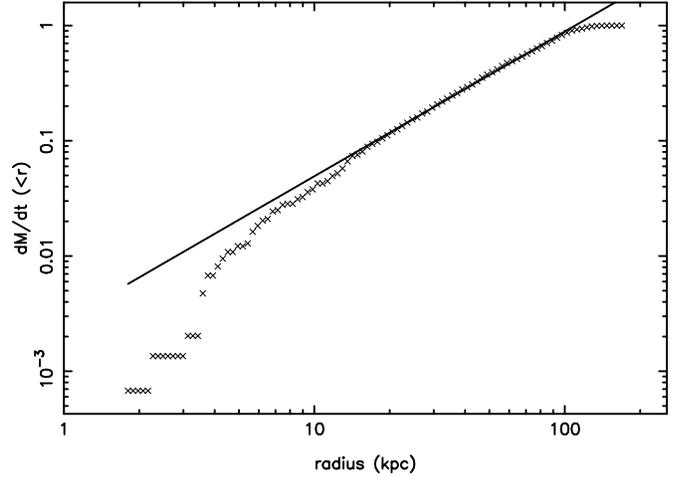,angle=270,width=8.7cm}
  \caption{Mass deposition profile produced by the multiphase code
  after 500 timesteps. Approximately 11\% of the gas has cooled out
  of the flow. The
  solid line represents a least squares fit to the profile between
  $15{\rm kpc} < r < 100{\rm kpc}$, and has a slope $1.26 \pm 0.01$.}
  \label{fig:mdot}
 \end{center}
\end{figure}
\begin{figure}
 \begin{center}
   \psfig{file=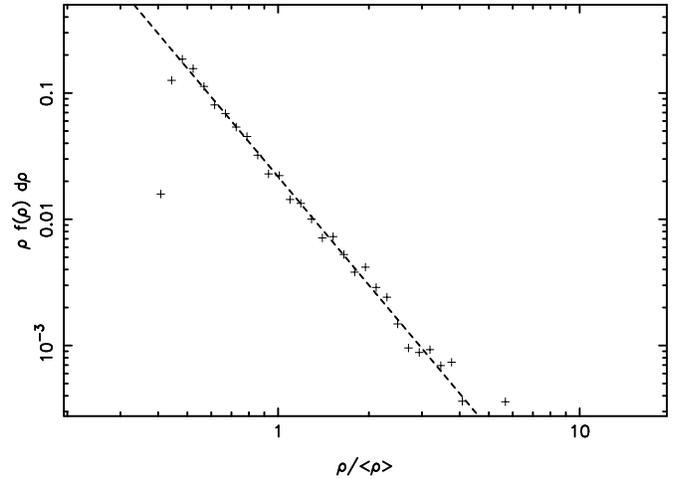,angle=270,width=8.7cm}
  \caption{Mass distribution function $\rho f {\rm d}\rho$ of the
  uncooled gas at radii $50 {\rm kpc} < r < 60 {\rm kpc}$ for the 
  multiphase code. We would expect the distribution to be unchanged from
  the initial distribution given by Equation~\ref{eq:cffuse}, with the 
  points on a power-law $\propto \rho^{-2.5}$. The dashed line 
  represents a least-squares fit (excluding the first two points), 
  and has a slope $-2.75 \pm 0.06$.}
  \label{fig:multif}
 \end{center}
\end{figure}
Figure~\ref{fig:mdot} shows the mass deposition profile $\dot{M}(<r)$,
produced by the multiphase method after 500 timesteps.  Roughly 11\%
of the gas has cooled from the flow.  The line marks a least-squares
fit to the data, with slope $\eta = 1.26 \pm 0.01$. The points within
$15$kpc of the centre of the flow have been excluded from this fit, as
there are too few particles here for the mass deposition profile to be
well sampled, as have particles with $r > 100{\rm kpc}$ which lie in
the corners of the cubical simulation volume.  The slope is slightly
steeper than the theoretical index of $\eta = 9/8$.
Figure~\ref{fig:multif} shows the mass distribution function $\rho f
{\rm d}\rho$ for particles at radii of between 50 and 60 kpc.  In
theory we would expect the distribution of phases given by
Equation~\ref{eq:cffuse} to remain unchanged with time, giving $\rho f
{\rm d}\rho \propto \rho^{-2.5}$.  The dashed line represents a
least-squares fit to the data, with gradient $-2.75 \pm 0.06$. This is
again steeper than we would expect. This is probably due to the phases
not comoving, an assumption made in N86. There is no condition in our
code to enforce this, and Equation~\ref{eq:eom2} implies that
high-density particles will receive a smaller pressure force than
low-density ones. The artificial viscosity will limit the degree to
which particles can interpenetrate, ensuring that the flow is largely
comoving, but it may not be sufficient to ensure that no slippage
occurs.  Applying a large bulk viscosity term in
Equation~\ref{eq:artvis} forces the particles to comove, and flattens
both slopes towards their theoretical values. However, this is not
suitable for general application as it degrades the shock capturing
ability of the method.
 
\begin{figure}
 \begin{center}
   \psfig{file=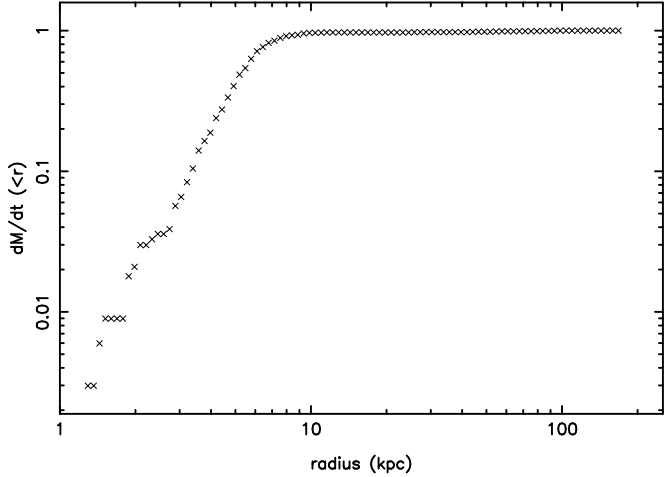,angle=270,width=8.7cm}
  \caption{Mass deposition profile produced by the standard 
  implementation of SPH after 500 timesteps. Approximately 5\% 
  of the gas has cooled out of the flow. Virtually all the matter
  has been deposited within the central $\sim 8$ kpc, as would 
  be expected of a single-phase cooling flow.}
  \label{fig:mdotsingle}
 \end{center}
\end{figure}
\begin{figure}
 \begin{center}
   \psfig{file=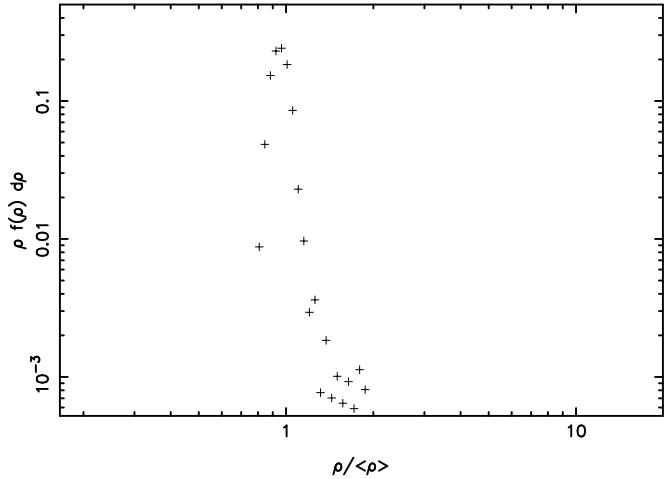,angle=270,width=8.7cm}
  \caption{Mass distribution function $\rho f {\rm d}\rho$ of the
  uncooled gas at radii $50 {\rm kpc} < r < 60 {\rm kpc}$ for the 
  standard implementation of SPH. All particles have densities close
  to the mean, and there is no trace of the original density 
  distribution.}
  \label{fig:singlef}
 \end{center}
\end{figure}
The mass deposition profile produced by the standard implementation of
SPH is shown in Figure~\ref{fig:mdotsingle}.  500 timesteps have
passed and this time about 5\% of the gas has cooled.  In contrast
with the multiphase results, this implementation produces a mass
deposition profile that is centrally concentrated, as would be
expected from a single-phase flow. This is supported by the mass
distribution function, which is plotted in Figure~\ref{fig:singlef}.
The gas has clearly evolved back to a single-phase state, with all
particles having a density close to the mean, and no trace of the
original density distribution remains.  In addition, the gradient of
the density profile has steepened from an initial value of $\rho
\propto r^{-0.75}$ (as given by Equation~\ref{eq:cfdens}) to $\rho
\propto r^{-1.15 \pm 0.03}$, close to the theoretical single-phase
gradient $\rho \propto r^{-1.2}$ (Thomas, 1988).

When compared to the standard implementation of SPH, it is clear that
our method performs well. Both the mass deposition profile and the
distribution of densities in the flow are close to the theoretical
values, whereas the standard implementation fails to reproduce the
properties of the flow correctly.

\subsection{Cosmological galaxy formation}
\label{sec:cosmic}
Our final test involves a simulation of the formation of a cluster of
galaxies. We use the initial conditions from the fiducial simulation
of Kay \etal (2000), who used it to test in detail the effect of
varying a number of numerical and physical parameters. The simulation
uses the standard cold dark matter (SCDM) cosmology, with $\Omega=1$,
$\Lambda=0$ and $h=0.5$\footnote{$H_0=100 h\, {\rm kms}^{-1}{\rm
    Mpc}^{-1}$}.  The baryon fraction was set from primordial
nucleosynthesis constraints, $\Omega_b h^2 = 0.015$ (Copi, Schramm \&
Turner 1995), and an unevolving gas metallicity of $0.5 Z_{\odot}$ was
used. The initial fluctuation amplitude was set so that the model
produces the same number density of rich clusters as is observed
today, with $\sigma_8$, the present-day linear rms fluctuation on a
scale of 8\hmpc, set to 0.6 (Eke, Cole \& Frenk 1996; Vianna \& Liddle
1996). Until $z \sim 1$ we adopt a comoving $\beta$-spline
gravitational softening equivalent to a Plummer softening of
$20$\hkpc, after which it is switched to a fixed physical softening of
$\sim \!10$ \hkpc. The minimum SPH resolution is set to match the
constraints imposed by the gravitational softening. $32^3$ particles
of both dark matter and gas were used, giving a mass per particle of
$8\times 10^9 h^{-1} M_{\odot}$ for the dark matter and $5\times 10^8
h^{-1} M_{\odot}$ for the gas. The simulation was started at $z \sim
24$ and was evolved to the present day.

For the purposes of this simulation, Equation~\ref{eq:coole} incorporates a source
term ${\cal H}$ to reflect the heating of gas due to a photoionising background. 
We assume that the background has a standard power-law form 
\begin{equation}
J(\nu) = J_{21}(z)\times10^{-21} \left(\frac{\nu}{\nu_{H_I}}\right) 
\hspace{0.2cm} {\rm ergs\, s}^{-1}{\rm cm}^{-2}{\rm sr}^{-1} {\rm Hz}^{-1},
\end{equation}
where 
\begin{equation}
J_{21}(z) = \frac{J_{21}^0}{1+(5/(1+z)^4)}
\end{equation}
is the flux at the H$_{\rm I}$ Lyman limit (Navarro \& Steinmetz,
1997); we take $J_{21}^0 = 1$ and $\alpha = 1$ here.  Photoheating is
implemented following Theuns \etal (1997).  The ultraviolet background
has the effect of imposing a minimum temperature on the gas, rapidly
heating it to $\sim 10^4K$ and ensuring that pressure gradients in the
gas remain shallow.  If the effects of photoionisation are not
included, then problems can occur when particles which have been in
free expansion since the start of the simulation, cooling to very low
temperatures with $T \propto (1+z)^2$, encounter the accretion shock
at the outer edges of halos.  This shock is generally poorly-resolved,
and neighbouring particles that have passed through the shock can make
an overwhelming contribution to the estimate of the pressure of the
cold particle, which can lead to the density being overestimated,
potentially by several orders of magnitude.  Incorporating
photoionisation serves as a simple way to limit this effect; an
alternative is to use the standard SPH estimate of the density for
calculating the emissivity in regions where the pressure gradient is
steep, as discussed in Section~\ref{sec:adiabatic}.

\begin{figure}
 \begin{center}
   \psfig{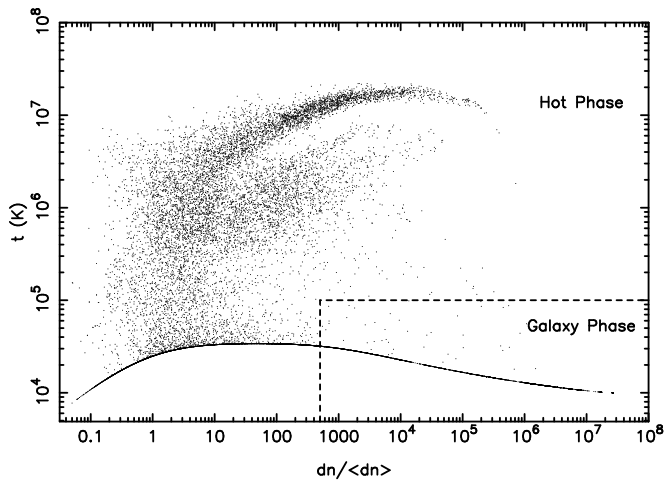}
   \caption{The temperature-density phase diagram of the gas at $z=0$ 
  in the multiphase cluster-formation simulation. Density is in
  units of the mean density. The diagram is divided into two phases;
  a \textit{galaxy} phase, containing gas which is
  overdense by at least a factor of $500$ and has a temperature of 
  $T \leq 10^5$K and a \textit{hot} phase, containing 
  all gas with a temperature of
  $T \geq 10^5$K and the gas which is not sufficiently
  overdense to be considered part of the galaxy phase.}
  \label{fig:mbarphase}
 \end{center}
\end{figure}
\begin{figure}
 \begin{center}
   \psfig{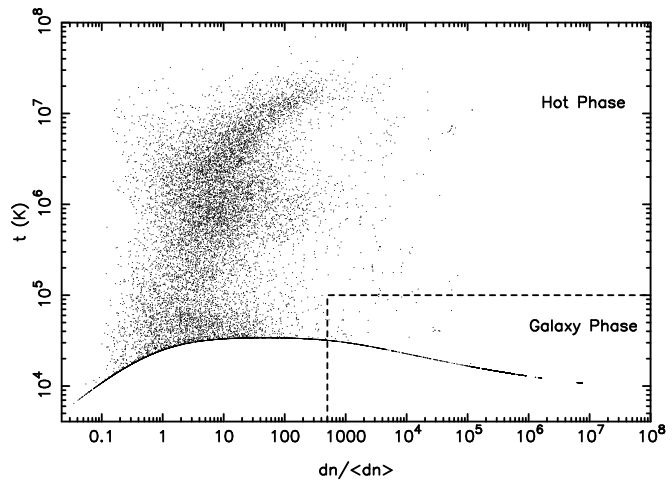}
  \caption{The temperature-density phase diagram of the gas at $z=0$ 
  in the standard SPH cluster-formation simulation. Density is in
  units of the mean density. The phases are defined in the same way
  as in Figure~\ref{fig:mbarphase}.}
  \label{fig:sbarphase}
 \end{center}
\end{figure}

Figures~\ref{fig:mbarphase} and~\ref{fig:sbarphase} illustrate the
temperature-density distribution of baryonic matter at $z=0$ produced
by the multiphase and standard codes.  We have divided the gas into
two phases; a \textit{galaxy} phase containing gas which is overdense
by at least a factor of $500$ and has a temperature of $10^3$K$\leq T
\leq 10^5$K and a \textit{hot} phase which includes all gas with a
temperature of $T \geq 10^5$K and the gas with $T \geq 10^3$K which is
not sufficiently overdense to be considered part of the galaxy phase.
Almost all the particles in the galaxy phase lie along a line with $T
\sim 1-2\times 10^4$K, which marks the point at which the radiative
cooling and photoheating rates balance. Most of these particles are in
the form of dense clumps with a size similar to the gravitational
softening length -- we refer to these clumps as galaxies. The
properties of these galaxies are calculating by first extracting all
particles within the galaxy phase from the simulation volume and then
running a 'friends-of-friends' group finder (Davis \etal 1985). The
most significant difference between the two simulations is in the mass
of the largest galaxy, which is nearly $50$\% more massive in the
single-phase simulation.  Kay \etal (2000) find that this is a result
of excessive cooling, as decoupling the hot and cold phases (Pearce
\etal 1999) reduces the final mass of the galaxy to a more reasonable
value.

\begin{figure}
 \begin{center}
   \psfig{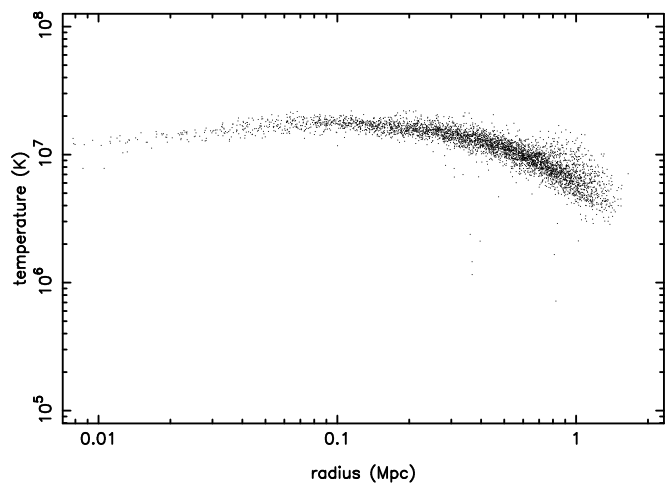}
  \caption{Radial temperature profile through the hot gas halo
  surrounding the largest galaxy found in the multiphase simulation.}
  \label{fig:mhalo}
 \end{center}
\end{figure}
\begin{figure}
 \begin{center}
   \psfig{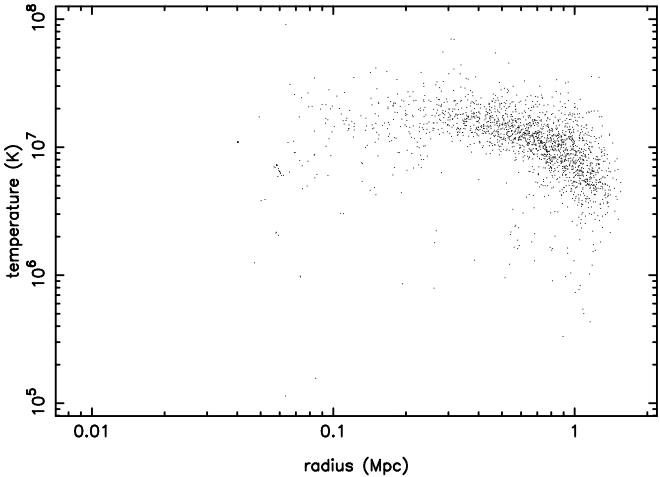}
  \caption{Radial temperature profile through the hot gas halo
  surrounding the largest galaxy found in the single phase simulation.}
  \label{fig:shalo}
 \end{center}
\end{figure}

The hot phase consists of particles that have been shock heated during
gravitational collapse, with the bulk kinetic energy of the gas being
converted to heat.  Figure~\ref{fig:mbarphase} shows a significant
quantity of hot ($T>10^7$K), high-density ($\rho/\langle\rho\rangle >
10^4$) gas which is not present in Figure~\ref{fig:sbarphase} but
\textit{is} seen in the single-phase simulations when radiative
cooling is not allowed.  This gas is located near the centre of large
dark matter halos, close to the central galaxy, and overestimation of
the already high gas density by the standard method leads to the gas
rapidly cooling and being accreted by the galaxy.  In contrast, the
multiphase method correctly estimates the density of the gas,
resulting in a slower cooling rate and the gas remaining at high
temperatures throughout the halo.  This effect can be seen in more
detail in Figures~\ref{fig:mhalo} and~\ref{fig:shalo}, which show the
radial temperature profile of the gas around the most massive galaxy
found in the simulation volume. The size of the galaxy is on the order
of the softening length, $\sim 10$\hkpc, and is not shown here.  The
results from the standard implementation clearly show two of the
problems inherent in the method. Firstly, particle temperatures drop
sharply within $r < 100$kpc - this is clear evidence for overcooling.
No such effect is visible in Figure~\ref{fig:mhalo} and the lack of
overcooling in our method is probably the principle reason for the
difference in mass of the largest objects in the two simulations.
Secondly, there is a complete absence of particles at radii $r <
50$kpc, whereas particles are found all the way in to $r \sim 10$kpc
in the multiphase simulations. This is an example of the effect
examined in Section~\ref{sec:twophase}, with particles close to the
central galaxy being forced away by an artificially asymmetric
pressure gradient in the single-phase method.

\section{DISCUSSION AND CONCLUSIONS}

\label{sec:disconc}

We have presented a multiphase implementation of Smoothed-Particle
Hydrodynamics (SPH), along with a number of tests to compare the 
performance of our method with standard implementations of SPH. 
The usual SPH formalism assumes that density is a smooth quantity,
varying negligibly on distances on the order of a typical smoothing
length. This is clearly not true in many situations in which SPH
is applied, such as simulations of galaxy formation, in which large
density contrasts are present. However, the pressure of the gas is 
expected to be a much smoother quantity because in almost all 
situations the sound-crossing time is shorter than the flow time 
across the smoothing sphere. We therefore summate the local pressure
at each particle, and calculate the density from the equation of state.

One situation in which our assumption will definitely not be true is
in the presence of shocks, in which density variations are generally
smaller than the variations in pressure. Sections~\ref{sec:shocktube}
and ~\ref{sec:adiabatic} demonstrate that our method handles shocks
acceptably, although the estimate of the density in the shock can be
inaccurate when strong shocks are resolved poorly.  This does not
alter the force calculation, which is dominated by the artificial
viscosity under such conditions, but is potentially significant when
radiative cooling is implemented, and it may be may be preferable to
use the standard estimate of the density for calculating the
emissivity in regions where the local pressure gradient is steep.  In
Section~\ref{sec:hupdate} we examine the degree to which steep density
gradients can be resolved by the two methods. Standard implementations
of SPH are shown to severely overestimate the density of particles
close to a region of high density, while these particles have their
density correctly estimated by our method.
In addition, we show in Section~\ref{sec:twophase} that unphysical 
forces can occur in the presence of steep density gradients, and while 
such forces are not completely eliminated by our method, 
they are greatly reduced. 
Section~\ref{sec:drag} examines the drag introduced by our implementation
of SPH. We find that at low Mach numbers drag is greatly reduced compared
to the SPH implementations of CTP95 and T98, while at higher velocities
the results are in broad agreement with the findings of TCP99, depending
mainly on the choice of smoothing kernel symmetrization. 

In Section~\ref{sec:cflow} we examine a simple spherically-symmetric,
constant pressure cooling flow. The fractional volume distribution $f
{\rm d}\rho$ of phases within the flow is taken to be a pure
power-law, which is shown by N86 to remain unchanged with time and
deposit mass as $\dot{M} (<r) \propto r^{9/8}$. The standard
implementation of SPH is shown to be unable to reproduce the expected
behaviour of this cooling flow, with conditions rapidly returning to a
single-phase state. Our code performs well, although the gradient of
both the mass distribution $\rho f {\rm d}\rho$ and the mass
deposition profile $\dot{M}$ are too steep. This is probably due to
phases not comoving in the flow, which is an assumption made in N86.
Application of a large bulk viscosity to force particles to comove
appears to reduce the problem, although this is not a suitable for
application generally as it results in the shock capturing ability of
the code being significantly degraded.

In Section~\ref{sec:cosmic} we examine the formation of a cluster of
galaxies.  The idealised problems examined in
Sections~\ref{sec:hupdate} and~\ref{sec:twophase} are shown to have
real analogues in the simulation using the standard SPH
implementation, with overcooling resulting from overestimating the
density of halo gas being clearly visible, and halo gas being forced
away from the central galaxy. No such effects are visible in the
simulation using our method. The lack of overcooling is also apparent
in the masses of the galaxies formed, with the largest galaxy being
nearly 50\% more massive in the single-phase simulation, in agreement
with the findings of Pearce \etal (1999), who found that decoupling
the galaxy from the hot halo gas produced a similar effect.  

Our method is an alternative to the standard formulation of SPH.  In
simulations without large density contrasts, the two give very similar
results.  However it represents a significant improvement over the
standard implementation of SPH when the gas component cannot be
assumed to be a single-phase fluid, such as galaxy formation and
cluster formation.  In addition, fully multiphase fluid flow can be
modelled, allowing SPH to be applied to simulations of cooling flows
and the intracluster medium.

\section*{Acknowledgments}

BWR acknowledges the support of a PPARC postgraduate studentship. PAT
is a PPARC Lecturer Fellow.  The authors would like to thank Rob
Thacker and Scott Kay for supplying initial conditions used in
testing, and Andy Fabian for useful suggestions.  We would also like
to thank the referee, James Wadsley, for helpful suggestions that have
greatly improved this paper. Parts of this work was conducted on the
SGI Origin platform using COSMOS Consortium facilities, funded by
HEFCE, PPARC and SGI.

\section*{References}

\paper{Balsara D.W.}{1995}{J.~Chem.~Phys.}{121}{357}
\paper{Bertschinger, E.}{1998}{\ARAA}{36}{599}
\paper{Bhattal A.S., Francis N., Watkins S.J., Whitworth A.P.}
  {1998}{\MN}{297}{435}
\paper{Blanton M., Cen R., Ostriker J.P., Strauss M.A.}{1999}
  {\ApJ}{522}{590}
\paper{Collela P., Woodward P.R.}{1984}{\JCP}{54}{174}
\paper{Copi C.J., Schramm D.N., Turner M.S.}{1995}{\ApJ}{455}{95}
\paper{Couchman H.M.P.}{1991}{\ApJ}{368}{23}
\paper{Couchman H.M.P., Thomas P.A., Pearce F.R.}{1995}
  {\ApJ}{452}{797} (CTP95)
\paper{Davis M., Efstatiou G., Frenk C.S., White S.D.M.}{1985}
  {\ApJ}{292}{371}
\paper{Edge A.C., Stewart G.C., Fabian A.C.}{1992}{\MN}{258}{177}
\paper{Eke V.R., Cole S., Frenk C.S.}{1996}{\MN}{282}{263}
\paper{Evrard A.E.}{1988}{\MN}{235}{911}
\paper{Fabian A.C.}{1994}{\ARAA}{32}{277}
\paper{Fabian A.C., Arnaud K.A., Nulsen P.E.J., Watson M.G.,
  Stewart G.C., McHardy I., Smith A., Cooke B., Elvis M.,
  Mushotzky R.F.}{1985}{\MN}{216}{923}
\paper{Flebbe O., M\"{u}nzel S., Herold H., Riffert H., Ruder H.}{1994}
  {\ApJ}{431}{754}
\paper{Frenk C.S., Evrard A.E., White S.D.M., Summers F.J.}{1996}
  {\ApJ}{472}{460}
\paper{Gingold R.A., Monaghan J.J.}{1977}{\MN}{181}{375}
\paper{Gingold R.A., Monaghan J.J.}{1983}{\MN}{204}{715}
\paper{Hawley J.F., Smarr L.L., Wilson J.R.}{1984}{\ApJ}{277}{296}
\paper{Hernquist L., Katz N.}{1989}{\ApJS}{70}{419} (HK89)
\press{Hutchings R.M., Thomas P.A.}{2000}{\MN} astro-ph/9903320
\paper{Kay S.T., Pearce F.R., Jenkins A., Frenk C.S.,
  White S.D.M., Thomas P.A., Couchman H.M.P.}{2000}{\MN}{316}{374}
\paper{Lucy L.B.}{1977}{\AJ}{82}{1013}
\paper{Monaghan J.J.}{1992}{\ARAA}{30}{543}
\paper{Monaghan J.J., Gingold R.A.}{1983}{\JCP}{52}{375}
\paper{Monaghan J.J.}{1997}{\JCP}{138}{801}
\paper{Navarro J., Steinmetz M.}{1997}{\ApJ}{478}{13}
\paper{Nelson R.P., Papaloizou J.C.B.}{1994}{\MN}{270}{1}
\paper{Nulsen P.E.J.}{1986}{\MN}{221}{377} (N86)
\conf{Nulsen P.E.J.}{1988}{NATO ASI Cooling flows
  in clusters and galaxies}{Fabian A.C.}{Kluwer, Dordrecht}{175}
\paper{Pearce F.R., Jenkins A., Frenk C.S., Colberg J.M.,
  Thomas P.A., Couchman H.M.P., White S.D.M., Efstathiou G.,
  Peacock J.A., Nelson A.H. (The Virgo Consortium)}{1999}{\ApJL}
  {521}{99}
\paper{Rasio F.A., Shapiro S.L.}{1991}{\ApJ}{377}{559}
\paper{Shapiro P.R., Martel H., Villumsen J.V., Owen J.M.}{1996}
  {\ApJS}{103}{269}
\paper{Sod G.A.}{1978}{\JCP}{27}{1}
\paper{Steinmetz M.}{1996}{\MN}{278}{1005}
\paper{Steinmetz M., M\"{u}ller E.}{1993}{\AaA}{268}{391}
\paper{Sutherland R.S., Dopita M.A.}{1993}{\ApJS}{88}{253}
\press{Thacker R., Tittley E.R., Pearce F.R., Couchman H.M.P.,
  Thomas P.A.}{1998}{\MN} astro-ph/9809221 (T98)  
\paper{Theuns T., Leonard A., Efstathiou G., Pearce F.R., Thomas P.A.}
{1998}{\MN}{301}{478}
\preprint{Tittley E.R., Couchman H.M.P., Pearce F.R.}{1999}
  astro-ph/9911017 (TCP99)
\conf{Thomas P.A.}{1988}{NATO ASI Cooling flows in clusters and galaxies}
  {Fabian A.C.}{Kluwer, Dordrecht}{361} 
\paper{Thomas P.A., Couchman H.M.P.}{1992}{\MN}{257}{11}  
\paper{Thomas P.A., Fabian A.C., Nulsen P.E.J.}{1987}{\MN}{228}{973}
  (TFN87)
\paper{Viana P.T.P., Liddle A.R.}{1996}{\MN}{281}{323}
\paper{White D.A., Jones C., Forman W.}{1997}{\MN}{292}{419}
\paper{Wood D.}{1981}{\MN}{194}{201}
\vfill

\end{document}